\documentclass[10pt,a4paper]{eucass}
\usepackage{subfigure}
\usepackage{hyperref}
\usepackage{graphicx}
\usepackage{color}
 \usepackage{epsfig}
\usepackage{amssymb}
\usepackage{amsmath}
\usepackage{amsthm}
\usepackage{booktabs}
\usepackage{latexsym}
\usepackage{eucal}
\usepackage{eufrak}
\usepackage{upref}
\usepackage{ulem}
\usepackage{lettrine}
\usepackage{verbatim}
\usepackage{array}
\usepackage{varwidth}

\begin{document}
\title{The role of presumed probability density function in the simulation of non premixed turbulent combustion}
\shorttitle{The role of PDF in non--premixed flames}
\author{A. Coclite$^{a,1}$, G. Pascazio$^{a,2}$, P. De Palma$^{a,3}$ and L. Cutrone$^{b,4}$}
\shortauthor{A. Coclite et al.}
\institution{$^{a}$Dipartimento di Meccanica, Matematica e Management (DMMM)\\ Centro di Eccellenza in Meccanica Computazionale (CEMEC)\\ Via Re David, 200, 70100, Bari (BA)-Italy\\ $^{b}$Centro Italiano di Ricerche Aerospaziali (CIRA)\\ Via Maiorise, 81043, Capua (CE)-Italy\\{\small $^1$a.coclite@poliba.it, $^2$pascazio@poliba.it, $^3$depalma@poliba.it, $^4$l.cutrone@cira.it}}
\maketitle

\begin{abstract}

Flamelet-Progress-Variable (FPV) combustion models~\cite{piercemoin2004} allow the evaluation of all thermo-chemical quantities in a reacting flow by computing only the mixture fraction $Z$ and a progress variable $C$. When using such a method to predict a turbulent combustion in conjunction with a turbulence model, a probability density function (PDF) is required to evaluate statistical averages (e.g., Favre average) of chemical quantities. The choice of the PDF is a compromise between computational costs and accuracy level. The aim of this paper is to investigate the influence of the PDF choice and its modeling aspects in the simulation of non-premixed turbulent combustion. Three different models are considered: the standard one, based on the choice of a $\beta$-distribution for $Z$ and a Dirac-distribution for $C$; a model employing a $\beta$-distribution for both $Z$ and $C$; a third model obtained using a $\beta$-distribution for $Z$ and the statistical most likely distribution (SMLD) for $C$ \cite{heinz}. The standard model, although widely used, doesn't take into account the interaction between turbulence and chemical kinetics as well as the dependence of the progress variable not only on its mean but also on its variance. The SMLD approach establishes a systematic framework to incorporate informations from an arbitrary number of moments~\cite{ihmea}, thus providing an improvement over conventionally employed presumed PDF closure models. The rational behind the choice of the three PDFs is described in some details and the prediction capability of the corresponding models is tested versus well-known test cases, namely, the Sandia flames~\cite{sandia}, and a test case for supersonic combustion provided by Cheng et al.~\cite{cheng}.\\
\end{abstract}

\section{Introduction}

The industrial and scientific communities are devoting major research efforts to identify and assess innovative technologies for  advanced propulsion system. Among such technologies, hydro-carbon combustion has been assumed as a key issue to achieve better propulsive performance and lower environmental impact. In order to improve the know-how to build more efficient engines with lower emissions it is necessary to enhance the knowledge of the combustion phenomena. In this context the simulation of turbulent reacting flows is very useful to cut down experimental costs and to achieve a thorough comprehension of the physical mechanisms involved. Turbulent combustion is a multi-scale problem, where the interaction between chemical kinetics, molecular, and turbulent transport occurs over a wide range of length and time scales. The numerical simulation of such phenomena with detailed chemistry is today prohibitive, so that a reduction model is often employed to condensate the reaction mechanisms and cut down the computational costs. Therefore, different approaches have been proposed to address this problem, such as the reduction of the chemical scheme in intrinsic low dimensional manifolds~(ILDM)~\cite{maas}; the flamelet-based approaches such as the flamelet-progress variable~(FPV)~\cite{pierce} or flame prolongation of ILDM~(FPI)~\cite{laminarhydrogen}; and Flamelet Generated Manifolds approach~(FGM)~\cite{oijen}. 

Our interest is devoted here to diffusive, either partially premixed or non-premixed, flames which constitute a specific class of combustion problems where fuel and oxidizer are not mixed before they enter into the combustion chamber. In this case mixing must bring reactants into the reaction zone so as to activate and maintain the combustion process. Non-premixed flames can be characterized by a local balance between diffusion and reaction~\cite{peters}and their structure can be described by a conserved scalar, the so-called mixture fraction. A diffusive flame can be viewed as an ensemble of thin locally one-dimensional structures embedded within the flow field. Each element of the flame front can then be described as a small laminar flame, also called {\it flamelet}. 
In this paper we focus on FPV approach for turbulent non-premixed flames. That relies on the use of only two degrees of freedom, namely, the mixture fraction and the progress variable, to map all of the thermo-chemical quantities involved in the process.
For the case of a turbulent flame one needs to define a probability density function (PDF) to compute the mean values of the thermo-chemical quantities. 
In particular, in the present work we are interested in the modeling of the PDF function required to evaluate chemical Favre averaged quantities. The definition of such a function is critical due to the poor knowledge of the two independent variables behaviour. The aim of this work is to provide an extension of the standard FPV model for turbulent combustion, applying the statistically most likely distribution (SMLD)~\cite{pope} approach to the progress variable PDF, maintaining a good compromise between computational costs and accuracy level. In the second section of this paper we present the rational for the definition of such a PDF. Then, three PDF models are considered and their role in the evaluation of non-premixed flames is analysed. In the third section the numerical results obtained in the simulation of the Sandia flames~\cite{sandia} and of a supersonic combustion~\cite{cheng} are discussed. The paper closes with summary and conclusions.\\

\section{Combustion model}

\subsection{The flamelet approach}

\begin{figure}
\begin{center}
\includegraphics[scale=0.4]{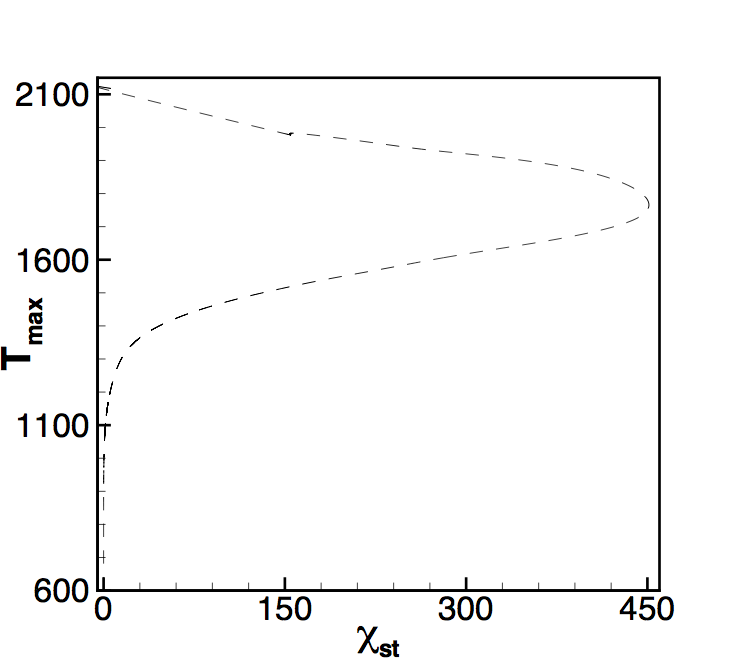}
\caption{S-shaped curve obtained plotting the maximum temperature of each flamelet versus the scalar stoichiometric dissipation rate.}
\label{scurve}
\end{center}
\end{figure}

The FPV model proposed by Pierce~\cite{pierce,piercemoin2004} is used in this work to evaluate all of the thermo-chemical quantities involved in the combustion process. This approach is based on the parametrization of the generic thermo-chemical quantities, $\phi$, in terms of two variables: the mixture fraction $Z$ and the progress variable $C$:
\begin{equation}
\label{phi}
{\phi=F_\phi(Z,C)}.
\end{equation}
Equation \eqref{phi} is taken as the solution of the steady laminar flamelet equation:
\begin{equation}
\label{slfe}
{-\rho\frac{\chi}{2}\frac{\partial^2 \phi}{\partial Z^2}=\dot\omega_\phi},
\end{equation}
where $\chi$ is the scalar dissipation rate modeled in terms of molecular diffusivity of mixture fraction $D_Z$, $\chi=2D_Z(\nabla Z)^2$; $\rho$ is the density; $\dot\omega_\phi$ is the source term related to $\phi$~\cite{pierce}. Each solution of equation~\eqref{slfe} is a flamelet and the solution variety over $\chi=\chi_{st}$, called S-curve, is shown in figure~\ref{scurve}. 
From equation \eqref{phi} one can obtain the Favre-averages of $\phi$ using the definitions:
\begin{equation}
\label{media}
{\widetilde\phi=\int\int F_\phi(Z,C)\widetilde{P}(Z,C)dZ dC},
\end{equation}
\begin{equation}
\label{varianza}
{\widetilde{\phi''^2}=\int\int (F_\phi(Z,C)-\widetilde\phi)^2\widetilde{P}(Z,C)dZdC},
\end{equation}
where $\widetilde P(Z,C)$ is the density-weighted PDF,
\begin{equation}
{\widetilde P(Z,C)=\frac{\rho P(Z,C)}{\overline{\rho}}},
\end{equation}
$P(Z,C)$ is the PDF and $\overline{\rho}$ is the Reynolds-averaged density. As usual, $\phi$ is decomposed as:
\begin{equation}
\phi=\widetilde \phi+ \phi'',
\qquad
\widetilde \phi = \frac{\overline{\rho \phi }}{\overline{\rho}}
\end{equation}
and
\begin{eqnarray}
\rho&=&\overline{\rho} +\rho',
\end{eqnarray} 
where $\phi''$ and $\rho'$ are the fluctuations.
This ensures that the filtering process does not alter the form of the conservation laws.\\
The choice of such a PDF plays a crucial role in the definition of the model, being a compromise between computational costs and accuracy level. In this respect, this paper provides an extension of the standard FPV turbulent combustion model combined with a Reynolds Averaged Navier-Stokes (RANS) equation solver~\cite{luigi}, where different fundamental hypotheses are used to define the PDF function for the progress variable, $C$. The influence of the different PDFs in the simulation of non-premixed turbulent combustion is the final aim of this research.
\subsection{Presumed probability density function modeling}
In order to investigate the role of the presumed PDF one can, first of all, use the Bayes theorem and take the PDF as the product between the marginal PDF of $Z$ and the conditional PDF of $C|Z$:
\begin{equation}
\label{bayes}
{\widetilde{P}(Z,C)=\widetilde P(Z)\widetilde P(C|Z)}.
\end{equation}
Therefore, one has to presume, or evaluate, the functional shape of such PDFs. Let us consider the marginal PDF, $\widetilde P(Z)$. It has been shown, by several authors, that the mixture fraction is best described by a passive scalar and that the PDF of a passive scalar can be approximated by a $\beta$~distribution function~\cite{cook,jimenez,wall}. The two parameter family of $\beta$-distribution in the interval $x\in [0,1]$ is given by:
\begin{equation}
\label{beta}
{P(x)=x^{a-1}(1-x)^{b-1}\frac{\Gamma(a+b)}{\Gamma(a)\Gamma(b)}},
\end{equation}
where $\Gamma(x)$ is the Euler function and $a$ and $b$ are two parameters related to $\widetilde x$ and $\widetilde{x''^2}$ respectively
\begin{equation}
\label{aeb}
{a=\frac{\widetilde x(\widetilde x- \widetilde{x}^2-\widetilde{x''^2})}{\widetilde{x''^2}}, \ \ \ b=\frac{(1-\widetilde x)(\widetilde x-\widetilde{x}^2-\widetilde{x''^2})}{\widetilde{x''^2}}}.
\end{equation}
For all the three models presented here, the $\beta$-distribution is employed for $\widetilde P(Z)$.\\
To presume the functional shape of the distribution of a reacting scalar, one needs to make some constitutive hypotheses. To simplify the problem, in this work we assume the statistical independence of $Z$ and $C$, so that, for all of the considered models, $\widetilde P(Z,C)=\widetilde P(Z)\widetilde P(C)$, namely $\widetilde{C}=\widetilde{C|Z}$. The most widely used hypothesis (model~A), implying great simplification in the theoretical framework, consists in assuming that the conditional PDF, $\widetilde P(C)$, can be modeled by a Dirac distribution. It can be shown that there is only one solution of equation \eqref{slfe} for each chemical state. With this criterion the Favre-average of a generic thermo-chemical quantity is given by:
\begin{equation}
\label{phidelta}
{\widetilde\phi=\int\int F_\phi(Z,C)\widetilde \beta(Z)\delta(C-\widetilde{C})dZdC=\int F_\phi(Z,\widetilde C)\widetilde \beta(Z)dZ}.
\end{equation} 
Therefore, one has only three additional transport equations (for $\widetilde Z$, $\widetilde{Z''^2}$ and $\widetilde C$) to evaluate all thermo-chemical quantities in the flow thus avoiding the expensive solution of a transport equation for each chemical species. However, it is well known that a reactive scalar~\cite{ihmea}, such as $C$, depends on a combination of solutions of equation~\eqref{slfe} for each chemical state and therefore its PDF cannot be accurately approximated by a Dirac distribution.\\
Therefore, the second model (model~B) is given by assuming that $Z$ and $C$ are distributed in the same way, namely, using a $\beta$~-~distribution, thus giving the following joint PDF:
\begin{equation}
\label{betabeta}
{\widetilde P(Z,C)=\widetilde\beta(Z)\widetilde\beta(C)}.
\end{equation}
This avoids the simplification seen before and, consequently, the model requires the evaluation of an additional transport equation for $\widetilde{C''^2}$. \\
Moreover, the probability distribution of a reacting scalar is often multi-modal, unlike the $\beta$ function, and its functional form depends on the turbulence-chemistry interaction. Therefore, one can think about a distribution built considering, as constraints, the only available informations, namely the value of $\widetilde Z$, $\widetilde{Z''^2}$, $\widetilde C$ and $\widetilde{C''^2}$. The third model (model~C) is obtained evaluating the conditional PDF as the statistically most likely distribution (SMLD)~\cite{ihmea}. It can be shown that if one knows only its first three moments, the PDF can be evaluated using ``Laplace's principle of insufficient reason''~\cite{pope}. The technique is developed following the statistical mechanics arguments presented by Heinz~\cite{heinz}. Relying on the knowledge of the first three moments of $\widetilde P(C)$, a unique measure, $S$, of the predictability of a thermodynamic state can be defined. $S$ is an entropy function depending on $\widetilde P(C)$, $S=S(\widetilde P(C))$~\cite{shannon} that can be thought of as the Boltzmann's entropy:
\begin{equation}
{S=-\int \widetilde P(C) \ln\Bigl(\frac{\widetilde P(C)}{Q(C)}\Bigr) dC},
\end{equation}
\noindent where $Q(C)$ is a bias density function to integrate information when no moments are known. In this paper the form of $Q(C)$ proposed by Pope~\cite{ihmea} is assumed. The goal is to construct a PDF that maximizes the entropy $S$. Following the Lagrangian optimization approach, the functional $S^*$ is defined by involving the constraints on the moments:
\begin{equation}
{S^*=-\int dC\Bigl\{ \widetilde P(C) \ln\Bigl(\frac{\widetilde P(C)}{Q(C)}\Bigr)+\sum_{n=1}^2 \mu_n C^n \widetilde P(C)-\frac{\widetilde P(C)}{Q(C)} \Bigr\}}.
\end{equation}
In the above equation $\mu_n$ are the Lagrange's multipliers while the last fraction term is introduced to normalize $\widetilde P(C)$. 
The expression for $\widetilde P(C)$, obtained evaluating the maximum of $S^*$, reads: 
\begin{equation}
\label{smld}
{\widetilde P(C)=\frac{1}{\mu_0}\exp\Bigl\{-\sum_{n=1}^2 \frac{\mu_n}{n}(C-\widetilde C)^n \Bigr\}},
\end{equation}
where:
\begin{eqnarray}
\mu_0&=&\int_0^1 dC \widetilde P(C),\\
-\mu_1&=&\int_0^1 dC \partial_C (\widetilde P(C))=\widetilde P(1)-\widetilde P(0),\\
1-\mu_2\widetilde{C''^2}&=&\int_0^1 dC \partial_C[(C-\widetilde C)\widetilde P(C)]=\widetilde P(1)-\widetilde C\mu_1.
\end{eqnarray}
since $Z$ and $C$ are bounded in $[0,1]$.

At this point the model still needs an additional assumption to be closed. Here we assume that the first and the last point of $\widetilde P(C)$ are equal to the first and last points of $\beta(C)$ evaluated with the given values of the mean and variance:
\begin{equation}
{\widetilde P(1;\widetilde C,\widetilde{C''^2})=\widetilde\beta(1;\widetilde C,\widetilde{C''^2})},\\
{\ \ \ \widetilde P(0;\widetilde C,\widetilde{C''^2})=\widetilde\beta(0;\widetilde C,\widetilde{C''^2})}.
\end{equation}
This assumption does not affect the multi-modal nature of the distribution, but simplifies the model implementation (there is no need to evaluate the roots of a non-linear system).
The major advantage of the SMLD approach over conventionally employed presumed PDF closure models is that it provides a systematic framework to incorporate an arbitrary number of moment information.
It is noteworthy that, since $C$ is used instead of $C|Z$ as argument of $\widetilde P$, also this model assumes statistical independence of $Z$ and $C$.

For the case of a turbulent flame, equation~\eqref{phi} must be written in terms of the Favre averages of $Z$ and $C$ and in terms of their variance. Using the model~A one can tabulate all chemical quantities in terms of $\widetilde Z$, $\widetilde{Z^{''2}}$ and $\widetilde C$ because of the properties of the $\delta$-distribution. On the other hand, models~B and C express $\phi$ in terms of $\widetilde{C^{''2}}$ too and therefore they need to evolve a transport equation also for $\widetilde{C^{''2}}$. 
Therefore,
in order to evaluate a diffusive flame with the three models described above, one has to define a transport equation for each of the flamelet variables, namely, $\widetilde Z$, $\widetilde{Z''^2}$, $\widetilde C$ and $\widetilde{C''^2}$. Only in the case of model~A, employing the $\delta$-distribution for $\widetilde{P}(C)$, the transport equation for $\widetilde{C''^2}$ is not needed and only three equations are solved. 
The transport equations read:
\begin{eqnarray}
\label{zmean}
 \partial_t(\overline{\rho}\widetilde{Z})+\vec\nabla\cdot(\overline{\rho}\widetilde{\vec u}\widetilde{Z})&=&
\vec\nabla\cdot\Bigl[\bigl( D+
D_{\widetilde{Z}}^t\bigr)\overline{\rho}
\vec\nabla\widetilde{Z}\Bigr],\\
\label{zvar}
\partial_t(\overline{\rho}\widetilde{Z''^2})+\vec\nabla\cdot(\overline{\rho}\widetilde{\vec u}\widetilde{Z''^2})&=&
\vec\nabla\cdot\Bigl[\bigl( D+D_{\widetilde{Z''^2}}^t\bigr)\overline{\rho}\vec\nabla\widetilde{Z''^2}\Bigr]-
\nonumber\\&-&\overline{\rho}\widetilde{\chi}+2\overline{\rho}D_{\widetilde Z}^t(\vec\nabla\widetilde{Z})^2,\\
\label{cmean}
\partial_t(\overline{\rho}\widetilde{C})+\vec\nabla\cdot(\overline{\rho}\widetilde{\vec u}\widetilde{C})&=&
\vec\nabla\cdot\Bigl[\bigl( D+D_{\widetilde{C}}^t\bigr)\overline{\rho}\vec\nabla\widetilde{C}\Bigr]+\overline{\rho}\overline{\dot\omega_C},\\
\label{cvar}
\partial_t(\overline{\rho}\widetilde{C''^2})+\vec\nabla\cdot(\overline{\rho}\widetilde{\vec u}\widetilde{C''^2})&=&
\vec\nabla\cdot\Bigl[\bigl( D+D_{\widetilde{C''^2}}^t\bigr)\overline{\rho}\vec\nabla\widetilde{C''^2}\Bigr]-
\nonumber\\&-&\overline{\rho}\widetilde{\chi}
+2\overline{\rho}D_{\widetilde C}^t(\vec\nabla\widetilde{C})^2+2\overline{\rho}\widetilde{C''\dot\omega''_C},
\end{eqnarray}
where $D$ is the diffusion coefficient for all of the species, given as $D=\nu/Pr$ evaluated assuming a unity Lewis number; $\nu$ is the cinematic viscosity and $Pr$ the Prandtl number; $D_{\widetilde Z}^t=D_{\widetilde{Z^{''2}}}^t=D_{\widetilde C}^t=D_{\widetilde{C''^2}}^t=\nu/Sc_{t}$ are the turbulent mass diffusion coefficients and $Sc_{t}$ the Shmidt turbulent number; $\dot{\omega}_{C}$ is the source for the progress variable. The gradient transport assumption for turbulent fluxes is used and the mean scalar dissipation rate, $\chi$ appear as a sink term in the equation \eqref{zvar}.\\
At every iteration, the values of the flamelet variables of the model are updated and the Favre-averaged thermo-chemical quantities are defined, using equation \eqref{media}. Such solutions provide the mean-mass-fractions which are used to evaluate the density, the enthalpy and all of the transport properties of the fluid. 


\section{Flow equations and numerical solution}

For an axisymmetric multi-component reacting compressible flow of $n$ species the RANS equations with k-w turbulence closure can be written as:
\begin{equation}
\label{floweq}
{\partial_t \vec Q+\partial_{x} (\vec E-\vec E_{\nu})+\partial_{r} (\vec F-\vec F_{\nu})= \vec S},
\end{equation} 
where $t$ is the time variable; $x$ and $r$ are the axial and the radial coordinate, respectively; $\vec Q$=($\overline \rho$,$\,\overline \rho \widetilde u_x$,$\,\overline \rho \widetilde u_r$,$\,\overline \rho \widetilde H$,$\,\overline \rho  k$, $\,\overline \rho \omega$,$\,\overline \rho \widetilde R_n$) is the 
vector of the conserved variables; $\overline \rho$, $(\widetilde u_x,\widetilde u_r)$, $\widetilde H$ indicate the Favre-averaged values of density, velocity components and specific total enthalpy, respectively; $k$ and $\omega$ are the turbulence kinetic energy and its specific dissipation rate; finally, $\widetilde R_n$ is a generic set of conserved variables related to the combustion model;
as described above, $\widetilde R_n$ is the set of independent variables of the flamelet model, namely, $\widetilde Z$, ${\widetilde{Z''^2}}$,
$\widetilde C$, ${\widetilde{C''^2}}$; $\vec E$, $\vec F$, and  $\vec E_v$, $\vec F_v$ are the inviscid and viscous flux vectors \cite{dsthesis}, respectively; $\vec S$ is the vector of the source term.\\
A cell-centered finite volume space discretization is used on a multi-block structured mesh. The convective and viscous terms are discretize by the third-order-accurate Steger and Warming~\cite{steger} flux vector-splitting-scheme and by second-order-accurate central differences, respectively. An implicit time marching procedure is used with a factorization based on the diagonalization procedure of Pulliamm and Chaussee~\cite{pulliam}, so as to allow a standard scalar alternating direction implicit (ADI) solution procedure~\cite{buelow97}. Only steady flows are dealt with in this paper. Therefore, the unsteady terms are eliminated from the governing equations and the ADI scheme is iterated in the pseudo-time until a residual drop of five orders of magnitude for all of the conservation-law equations~\eqref{floweq} has been achieved.  
Additional details, when needed, are provided in the following for each single application.

\section{Numerical results}

This section provides the comparison among the results obtained using the three combustion models so as to assess the influence that the PDF choice may have in the prediction of turbulent non-premixed flames. The well-knonw subsonic Sandia flames~\cite{sandia} are considered at first, then a supersonic test case is analysed, whose experimental data are available in the literature~\cite{cheng}. The steady flamelet evaluations, for both test cases, have been performed using the FlameMaster code~\cite{flamemaster}.\\
\subsection{Sandia flames test case}
\begin{figure}
\begin{center}
\includegraphics[scale=0.6]{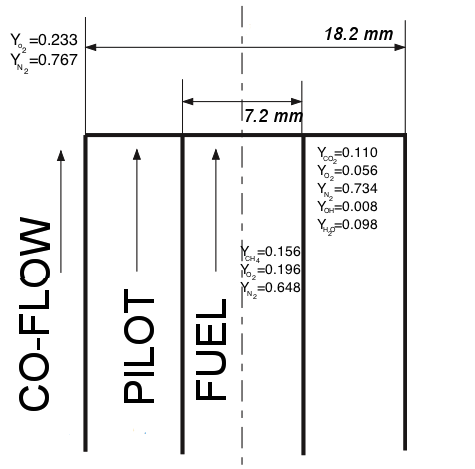}
\caption{Schematic of the Sandia Flame burner.}
\label{sandia_geom}
\end{center}
\end{figure}
\begin{figure}[!h]
\begin{center}
\includegraphics[scale=0.5]{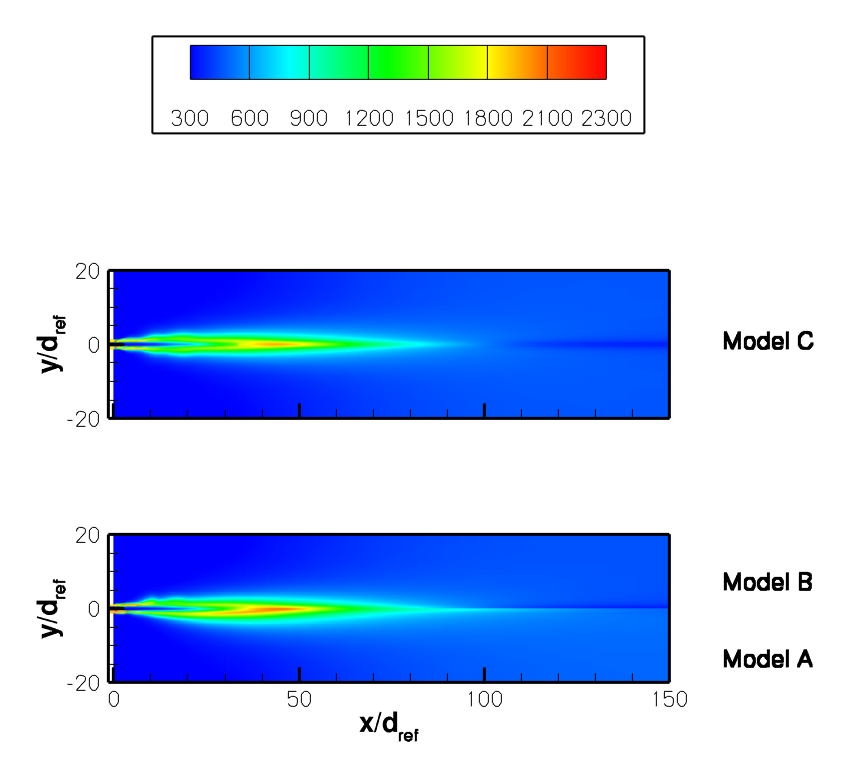}
\caption{Temperature contours for the Sandia Flame E: model A, model B and model C.}
\label{E_temp}
\end{center}
\end{figure}
\begin{figure}[!h]
\begin{center}
\includegraphics[scale=0.5]{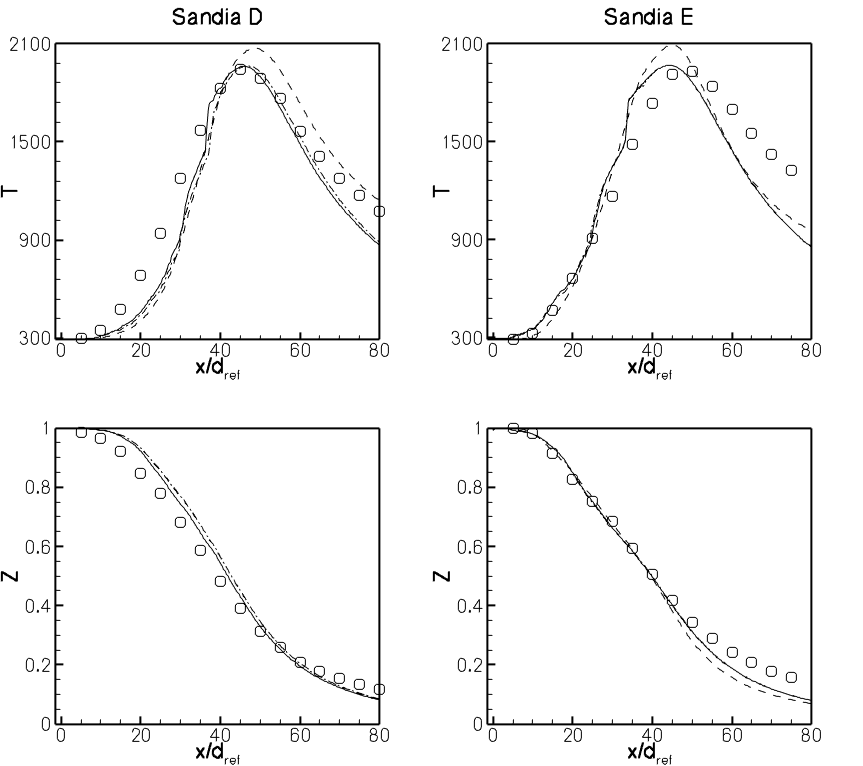}
\caption{Temperature and mixture fraction distributions along the axis: Flame D (left), Flame E (right). Model A, dashed line; model B, dashed-dotted line; model C, solid line; experimental data, symbols.}
\label{center}
\end{center}
\end{figure}
\begin{figure}
\begin{center}
\includegraphics[scale=0.5]{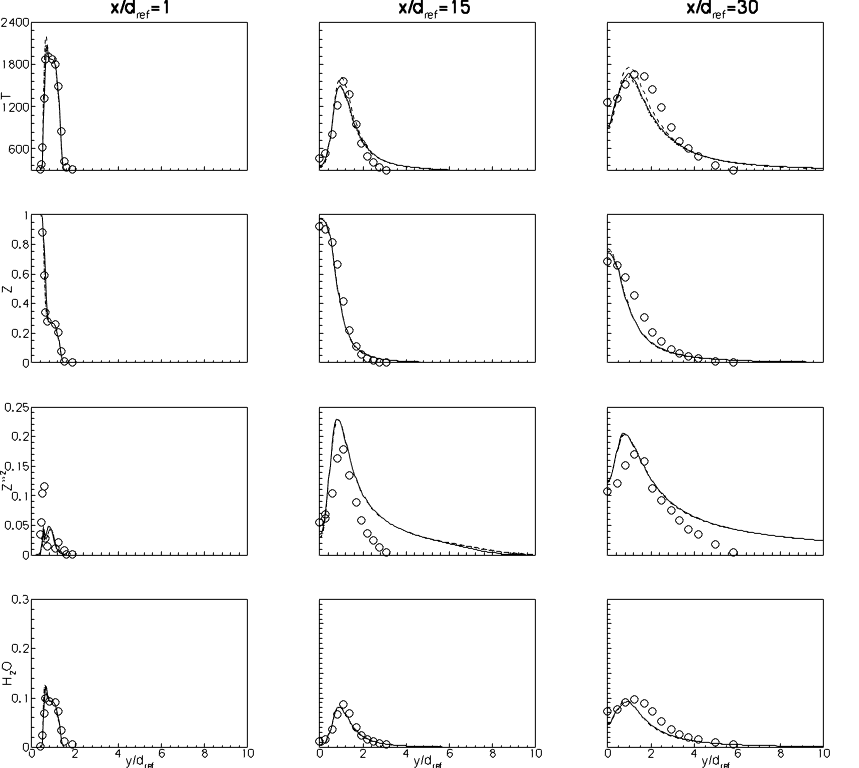}
\caption{Thermodynamic quantities distributions in Sandia Flame D along several sections taken at $x/d_{ref}=1$, $x/d_{ref}=15$, and $x/d_{ref}=30$  from the burner.
Model A, dashed line; model B, dashed-dotted line; model C, solid line; experimental data, symbols.}
\label{flame_D}
\end{center}
\end{figure}
\begin{figure}
\begin{center}
\includegraphics[scale=0.7]{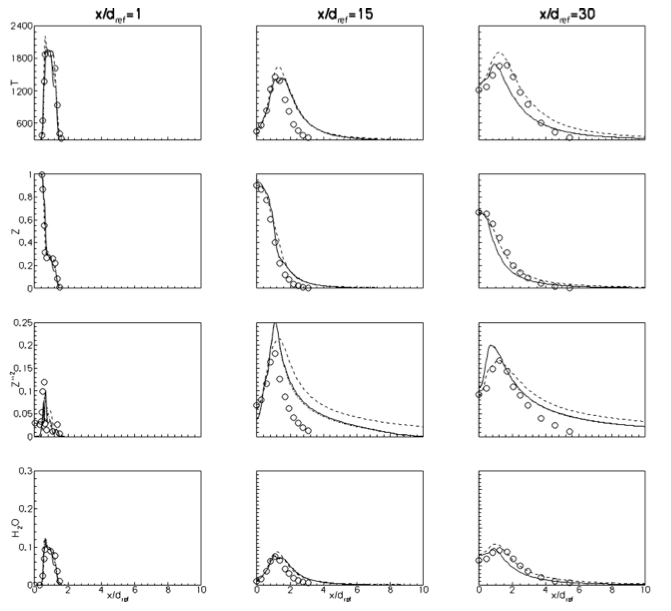}
\caption{Thermodynamic quantities distributions in Sandia Flame E along several sections taken at $x/d_{ref}=1$, $x/d_{ref}=15$, and $x/d_{ref}=30$  from the burner.
Model A, dashed line; model B, dashed-dotted line; model C, solid line; experimental data, symbols.}
\label{flame_E}
\end{center}
\end{figure}
The Sandia Flames are two piloted partially premixed methane-air diffusion flames burning at pressure equal to 
$100.6$ kPa and at different Reynolds numbers, $Re$, based on the nozzle diameter, the jet bulk velocity, and the kinematic 
viscosity of the fuel. 
The diameter of the nozzle of the central jet is $d_{ref}=7.2$ mm and the internal and external diameters of the annular pilot 
nozzle are equal to  $7.7$ mm and $18.2$ mm, respectively. 
The fluid jet is a mixture of $75\%$ air and $25\%$ methane by volume. Partial pre-mixing with air reduces the 
flame length and produces more robust flame than pure CH$_{4}$~\cite{sandia}. Consequently, the flame may be operated at high $Re$
with little effect of local extinction even with a little pilot. Flame D ($Re=22400$) presents very low degree of local extinction, whereas Flame E ($Re =33600$) has significant and increasing probability of local extinction near the pilot. 
The pilot is a mixture of air with the main methane combustion products, namely C$_2$H$_2$, H$_2$, CO$_2$ and N$_2$, with the same 
enthalpy at the equivalence ratio $\Phi =0.77$ corresponding to the equilibrium composition $\widetilde Z=0.27,\ \widetilde{Z''^2}=0.0075,\ \widetilde C=1,\ \widetilde{C''^2}=0$. 
The oxidizer air (Y$_{O_2}$=0.233, Y$_{N_2}$=0.767) is supplied as a co-flow at $291$ K. The schematic of the Sandia Flame burner is shown in figure~\ref{sandia_geom}.

The computational domain is axisymmetric and includes a part of the burner; it has a length of $150\ d_{ref}$ and $27\ d_{ref}$ along the axial 
and radial directions, respectively, and has been discretized using about $45000$ cells. Computations have been carried out using the 
combustion scheme described by the GRI-MECH~3.0~\cite{grimech30}: $325$ sub-reactions upon $53$ species. The flamelet library is computed 
over a grid with $125$ uniformly distributed points in the $\widetilde Z$ and $\widetilde C$ directions and $25$ uniformly distributed points
in the $\widetilde{Z''^2}$ and $\widetilde{C''^2}$ directions. Indeed, when considering model~A the grid is built considering only 
$\widetilde Z$, $\widetilde{Z''^2}$ and $\widetilde C$.

The temperature field obtained by the three combustion models for the case of the Sandia Flame E is presented in figure~\ref{E_temp}. 
Moreover, figure~\ref{center} provides the temperature distributions along the axial direction. It appears that in the near-burner region 
model~C is in better agreement with the experimental data than the other two models. Moving away from the burner ($x>20\ d_{ref}$) 
the agreement deteriorates; this is probably be due to the accuracy limits of the RANS approach in the prediction of the mixing process 
that greatly affects combustion.
From this two set of figures one can see that there is an improvement provided by model~B and model~C in the evaluation of the flame core and of the flame shapes.\\
The radial distributions of temperature, mixture fraction, mixture fraction variance and $H_{2}O$ mass fraction 
at several sections taken at the axial coordinate $x/d_{ref}$ equal to 1, 15, and 30 are shown
in figures \ref{flame_D} and \ref{flame_E} for the Sandia Flame D and Sandia Flame E, respectively.
One can see that the results of model B (corresponding to presume that $\widetilde P(Z,C)=\beta(Z)\beta(C)$) and model C 
(corresponding to the choice $\widetilde P(Z,C)=\beta(Z)P_{SML,2}(C)$) are in better agreement with the experimental data~\cite{sandia} 
than the results of model A (corresponding to the standard choice $\widetilde P(Z,C)=\beta(Z)\delta(C-\widetilde{C|Z})$). 
The improvement in the prediction of the flame core is evident in the figure~\ref{center}, where the differences in the results by the three models are more relevant. 
Envisaging a model that does not require the hypothesis of statistically independence of $Z$ 
and $C$ should considerably enhance the prediction accuracy also in the regions close to the burner.

\subsection{Supersonic combustion test case}

\begin{figure}
\begin{center}
\includegraphics[scale=0.6]{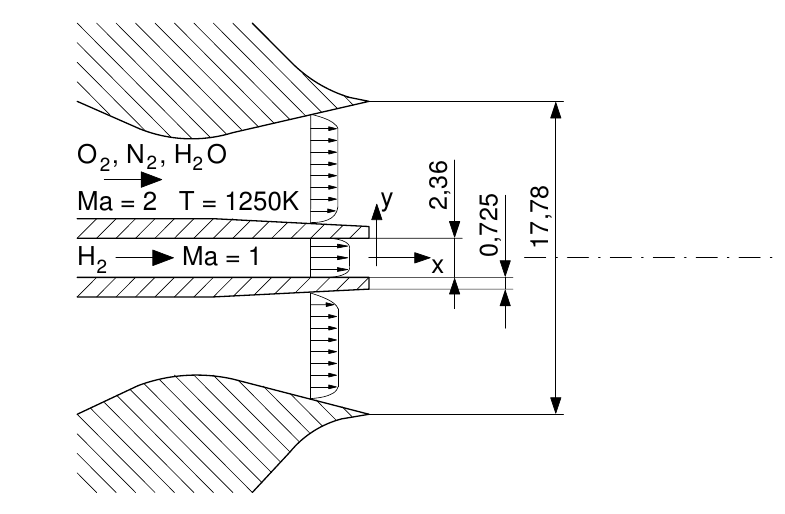}
\caption{Schematic of the Cheng's burner.}
\label{cheng_geom}
\end{center}
\end{figure}
The test case is the hydrogen-air supersonic combustion set-up proposed by Cheng et al.~\cite{cheng}. 
Figure~\ref{cheng_geom} shows the supersonic burner that provides an annular, axisymmetric jet of hot, vitiated wet air at Mach number equal to $2$, average axial velocity 
of $1417\ m/s$, temperature of $1250\ K$ and pressure of $107\ kPa$. The vitiated air is composed of the following set of mass fractions: 
$Y_{O_2}=0.245$, $Y_{H_2O}=0.175$ and $Y_{N_2}=0.58$. The exit conditions of the air stream are given by a pre-combustion at low temperature~\cite{cheng}. The hydrogen exit is estimated as a chocked flow, average axial velocity of $1780\ m/s$, temperature of $545\ K$ 
and pressure of $112\ kPa$. The diameter of the fuel stream is $d_{ref}=2.362\ mm$, taken as the reference length. At the inlet section, the values of the inner and outer diameters of the vitiated air stream are equal to 3.812 mm and 17.78 mm, respectively~\cite{jarret}. 
The computational domain is axisymmetric and includes the divergent part of the air nozzle; it extends $150\ d_{ref}$ and $50\ d_{ref}$ along the axial and radial directions, respectively, and has been discretized using about $100000$ cells. 
The evaluation of the flamelet library has been performed using the Vajda kinetic scheme~\cite{vajda}: $19$ sub-reactions upon $7$ species. The flamelet library has been computed over a grid with $250$, uniformly distributed points, 
in the $\widetilde Z$ and $\widetilde C$ directions, and $50$ uniformly distributed points in the $\widetilde{Z''^2}$ and $\widetilde{C''^2}$ directions.
\begin{figure}
\begin{center}
\includegraphics[scale=0.37]{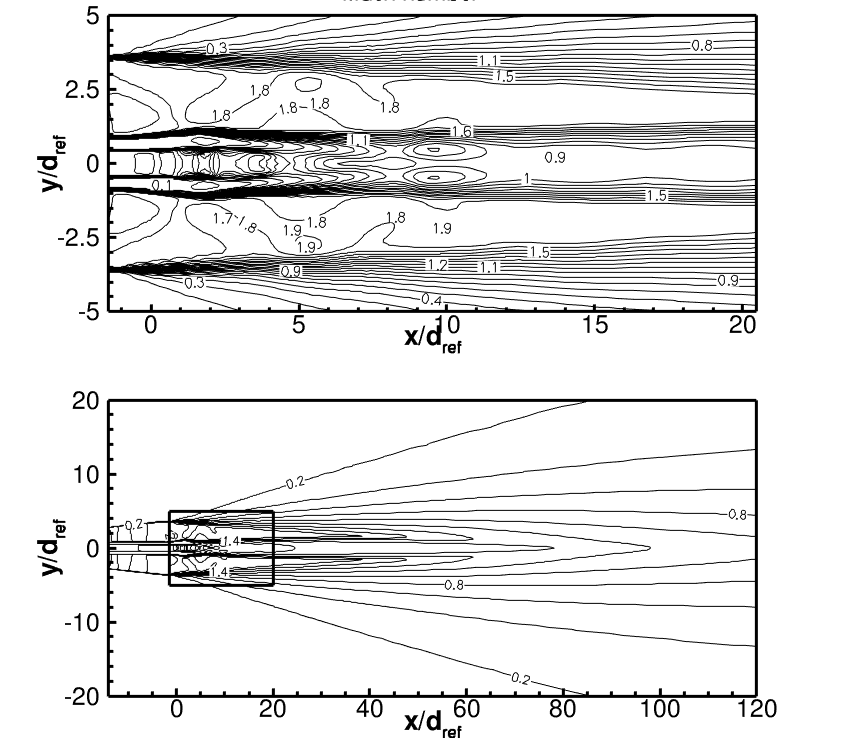}
\caption{Mach's number contours with a close up of the near-burner region (top) of the Cheng's Flame combustion chamber~\cite{cheng}.}
\label{cheng_mach}
\end{center}
\end{figure}
\begin{figure}
\begin{center}
\includegraphics[scale=0.35]{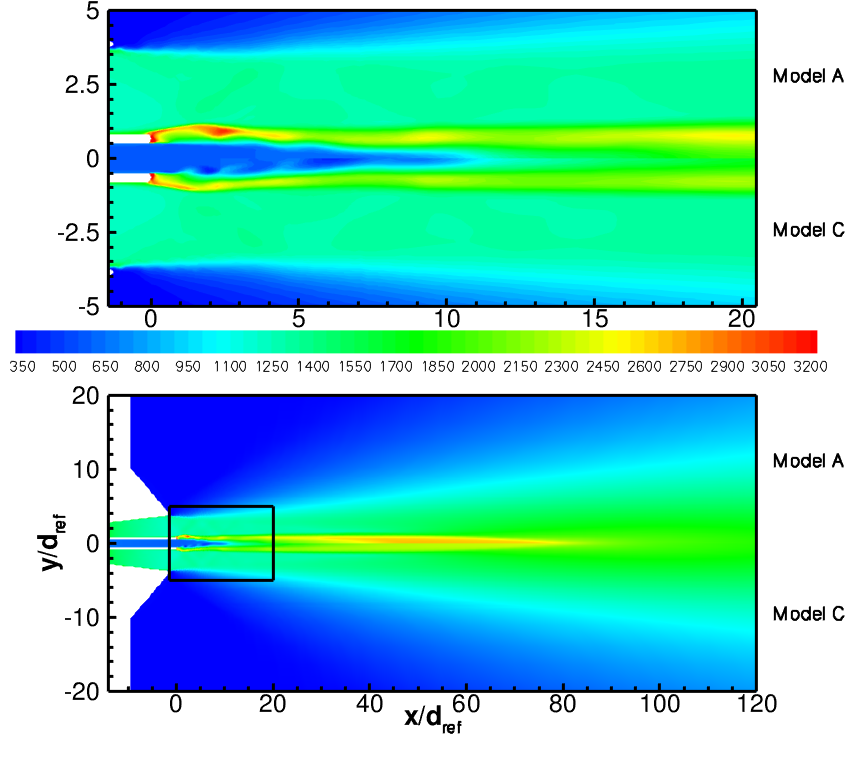}
\caption{Temperature contours with a close up of the near-burner region (top) of the Cheng's combustion chamber~\cite{cheng}.}
\label{cheng_temp}
\end{center}
\end{figure}
\begin{figure}
\begin{center}
\includegraphics[scale=0.5]{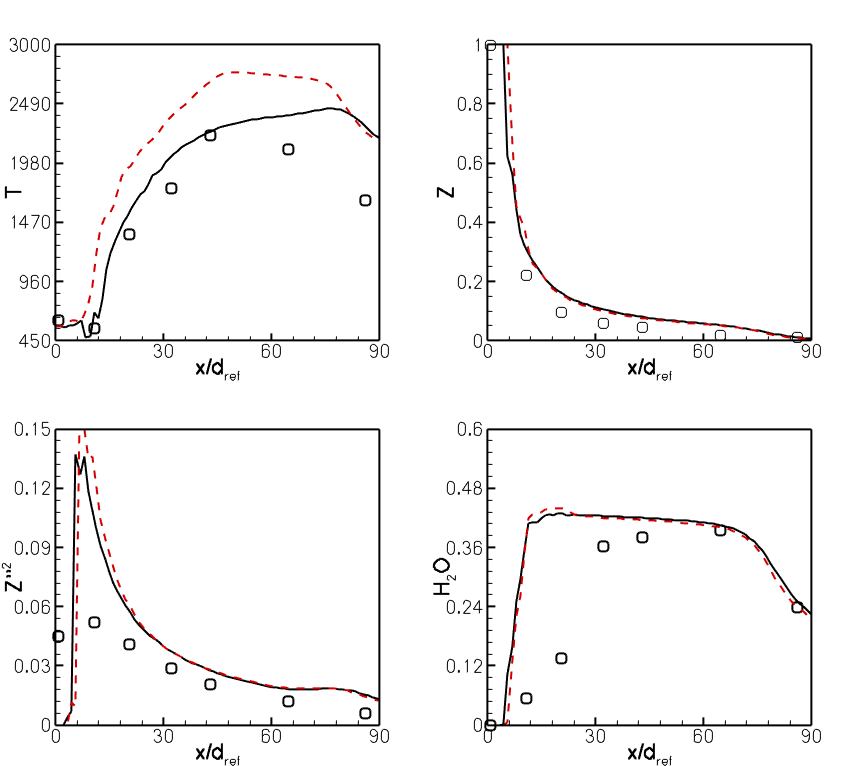}
\caption{Cheng combustion chamber thermo-chemical distributions along the axis~$(y/d_{ref}=0)$. Model~A red dashed line; Model~C, solid black line; Symbols, experimental data~\cite{cheng}.}
\label{cheng_axial}
\end{center}
\end{figure}
\begin{figure}
\begin{center}
\includegraphics[scale=0.56]{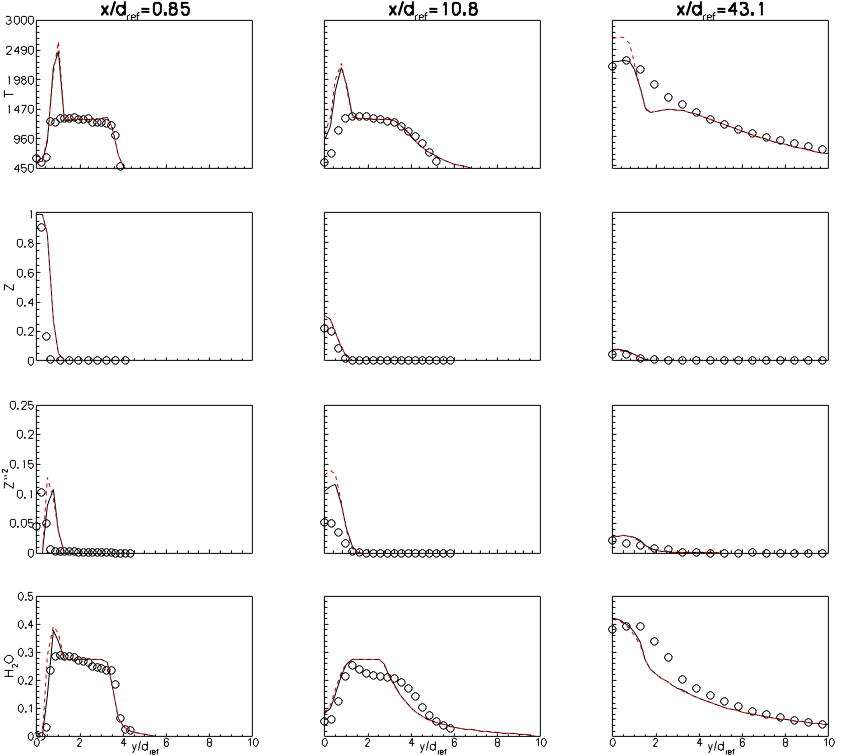}
\caption{Cheng combustion chamber thermo-chemical distributions taken at $x/d_{ref}=0.85$, $x/d_{ref}=10.8$, $x/d_{ref}=33.1$, and $x/d_{ref}=86.1$. Model~A, solid line;  Model~C, solid black line; Symbols, experimental data~\cite{cheng}.}
\label{cheng_radial}
\end{center}
\end{figure}

\noindent Figure~\ref{cheng_mach} provides the Mach number contours; one can see, through the close up of the near-burner region, that the initial velocity conditions are fulfilled.
In figure~\ref{cheng_temp}, a qualitative comparison between the temperature distribution evaluated with the model~A and the model~C, here one can see that the two predictions of flame shape are very different. Model~A evaluate a reaction zone attached to the burner longer than that evaluated with model~C; the first one extends up to about $4\ d_{ref}$ and the second one extends up to about $2\ d_{ref}$. 
Figure~\ref{cheng_axial} shows the main thermodynamic quantities taken on the axis of the burner, it is clear that the flame core is not well reproduced using model~A. This is probably due to the PDF modeling since it is proved~\cite{gerlinger} that the Vajda kinetic scheme can well predict such a flame. The differences between the prediction made by the two model are relevant especially in the temperature evaluation. 
However, both models provide an high temperature stripe in the near axis region, the flame is being attached to the burner.
This is probably due to the fact that FPV models are built whit the low-mach-number hypothesis~\cite{supersonic_pitsch}. In particular model~A evaluates that the reactions may occur in the entire region in which the two flows ($H_{2}$--wet-air) come up against each other, providing a fast reaction rate for the combustion process. 
Figure~\ref{cheng_radial} shows the distribution of the main thermo-chemical quantities taken at several radial sections. Both models provide results in a good agreement with experimental data~\cite{cheng} for $y>d_{ref}$. Figures~\ref{cheng_axial} and~\ref{cheng_radial} shows that both the flame length and the mixture fraction consumption are well predicted with the model~C.
All the results are in agreement with the state of the art of FPV models combined with RANS turbulent modeling~\cite{gerlinger, supersonic_pitsch, gao}.
 
\section*{Conclusions}

This paper provides an extension of standard FPV model combined with a RANS solver introducing the SMLD approach to describe the progress variable distribution. The work is composed of four sections. The first one is an introduction to the problem of the modeling PDF in non--premixed combustion. The second section describes the developed combustion model developed with a new closure method for the SMLD technique. The third section provides the flow governing equations, the additional transport equations of the combustion models, and numerical method. The results are discussed in the last section for the case of a subsonic $CH_4$--air flame and a $H_2$--air Mach $2$ flame. The analysis is performed in order to validate the applicability of the developed models. In the first case, the flow is very simple and the steady flamelet equation well represents the phenomenon; in the second case, although some of the assumptions of the FPV model are not fulfilled, the numerical
results can be considered satisfactory.  

%
\newcommand{\auth}{\textsc}
\newcommand{\tit}{\textrm}
\newcommand{\jou}{\textit}
\bibliographystyle{elsarticle-num.bst}
\bibliography{biblio.bib}

\begin{thebibliography}{10}
\expandafter\ifx\csname url\endcsname\relax
  \def\url#1{\texttt{#1}}\fi
\expandafter\ifx\csname urlprefix\endcsname\relax\def\urlprefix{URL }\fi
\expandafter\ifx\csname href\endcsname\relax
  \def\href#1#2{#2} \def\path#1{#1}\fi

\bibitem{piercemoin2004}
C.~D. Pierce, P.~Moin, Progress-variable approach for large-eddy simulation of
  non-premixed turbulent combustion, J. Fluid Mech. 504 (2004) 73--97.

\bibitem{heinz}
S.~Heinz, Statistical mechanics of turbulent flows, Springer-Verlag, 2003.

\bibitem{ihmea}
M.~Ihme, H.~Pitsch, Prediction of extinction and re-ignition in non-premixed
  turbulent flames using a flamelet progress variable model. 1 {A} priori study
  and presumed {PDF}, Combust. Flame 155 (2008) 70--89.

\bibitem{sandia}
Sandia National Laboratories, TNF Workshop,
  \texttt{http://www.ca.sandia.gov/TNF}.

\bibitem{cheng}
T.~S. Cheng, J.~A. Wehrmeyer, R.~W. Pitz, Simultaneous temperature and
  multispecies measurement in a lifted hydrogen diffusion flame, Combust. Flame
  91 (1992) 323--345.

\bibitem{maas}
U.~Maas, S.~Pope, Simplifying chemical kinetics- intrinsic low-dimensional
  manifolds in composition space, Combust. Flame 88 (1992) 239--264.

\bibitem{pierce}
C.~Pierce, Progress-variable approach for large-eddy simulation of turbulent
  combustion, {PhD Thesis}, Stanford University (2001).

\bibitem{laminarhydrogen}
O.~Gicquel, N.~Darabiha, D.~Thevenin, Laminar premixed hydrogen/air counterflow
  flame simulations using flame prolongation of ildm with differential
  diffusion, Proc. Combust. Inst. 28 (2000) 1901--1908.

\bibitem{oijen}
J.~V. Oijen, L.~D. Goey, Modelling of premixed laminar flames using
  flamelet-generated manifolds, Combust. Sci. Technol. 161 (2000) 113--137.

\bibitem{peters}
N.~Peters, Turbulent combustion, Cambridge University Press, 2000.

\bibitem{pope}
S.~Pope, {PDF} methods for turbulent reactive flows, Prog. Energy Combust. Sci.
  11 (1985) 119--192.

\bibitem{luigi}
L.~Cutrone, P.~{De Palma}, G.~Pascazio, M.~Napolitano, A {RANS}
  flamelet-progress-variable method for computing reacting flows of real-gas
  mixtures, Comput. Fluids 39 (2010) 485--498.

\bibitem{cook}
A.~W. Cook, J.~J. Riley, A subgrid model for equilibrium chemistry in turbulent
  flows, Phys. Fluids 6 (1994) 2868--2870.

\bibitem{jimenez}
J.~Jimenez, A.~Linan, M.~M. Rogers, F.~J. Higuera, A priori testing of subgrid
  models for chemically reacting non-premixed turbulent shear flows, J. Fluid
  Mech. 349 (1997) 149--171.

\bibitem{wall}
C.~Wall, B.~J. Boersma, P.~Moin, An evaluation of the assumed beta probability
  density function subgrid-scale model for large eddy simulation of
  non-premixed, turbulent combustion with heat release, Phys. Fluids 12 (2000)
  2522--2529.

\bibitem{shannon}
C.~H. Shannon, A mathematical theory of communication, Bell system technical
  journal 27 (1948) 379,423.

\bibitem{dsthesis}
D.~A. Schwer, {Numerical study of unsteadiness in non-reacting and reacting
  mixing layers}, {PhD Thesis}, The Pennsylvania State University (1999).

\bibitem{steger}
J.~L. Steger, R.~F. Warming, {Flux vector splitting of the inviscid gas-dynamic
  equations with applications to finite difference methods}, J. Comput. Phys.
  40 (1981) 263--293.

\bibitem{pulliam}
T.~H. Pulliam, D.~S. Chaussee, A diagonal form of an implicit factorization
  algorithm, J. Comput. Phys. 39 (1981) 347--363.

\bibitem{buelow97}
P.~E.~O. Buelow, D.~A. Schwer, J.-Z. Feng, C.~L. Merkle, D.~Choi, {A}
  preconditioned dual time diagonalized {ADI} scheme for unsteady computations,
  in: {A}IAA {P}roceedings, 1997.

\bibitem{flamemaster}
H.~Pitsch, Flamemaster v3.3. a c++ computer program for 0d combustion and 1d
  laminar flame calculationsAvailable at \texttt{http://www.stanford.edu/$\sim$
  hpitsch}.

\bibitem{grimech30}
G.~P. Smith, D.~M. Golden, M.~Frenklach, N.~W. Moriarty, B.~Eiteneer,
  M.~Goldenberg, C.~T. Bowman, R.~K. Hanson, S.~Song, W.~C. Gardiner, V.~V.
  Lissianski, Z.~Qin, 2000, \texttt{http://www.me.berkeley.edu/gri\_mech/}.

\bibitem{jarret}
O.~Jarret, A.~D. Cutler, R.~R. Antcliff, J.~A. Wang, Twenty-fifth {JANAF}
  combustion meeting, {CPIA} publications 498.

\bibitem{vajda}
S.~Vajda, H.~Rabitz, A.~Yetter, Effects of thermal coupling and diffusion on
  the mechanism of h2 oxidation in steady premixed laminar flames, Combustion
  and Flame 82 (1990) 270--297.

\bibitem{gerlinger}
P.~Gerlinger, Study of a multi-variate-beta-{PDF} for species distributions and
  model improvement, Tech. rep. (2007).

\bibitem{supersonic_pitsch}
V.~E. Terrapon, F.~Ham, R.~Pecnik, H.~Pitsch, A flamelet-based model for
  supersonic combustion, Tech. rep. (2009).

\bibitem{gao}
Z.~Gao, C.~Lee, A flamelet model for turbulent diffusion combustion in
  supersonic flow, Science China Tecnological Sciences 53~(12) (2010)
  3379--3388.

\end{thebibliography}
\end{document}